\documentclass[usenatbib]{mn2e}
\usepackage{graphicx}
\usepackage{lscape}
\usepackage{amssymb}
\usepackage{color}
\usepackage{float}
\usepackage{amsmath}
\usepackage[title]{appendix}
\usepackage{booktabs}
\usepackage{multirow}
\usepackage{siunitx}

\newcommand{\bn}{\begin{enumerate}}
\newcommand{\en}{\end{enumerate}}
\newcommand{\bi}{\begin{itemize}}
\newcommand{\ei}{\end{itemize}}

\def\gtorder{\mathrel{\raise.3ex\hbox{$>$}\mkern-14mu
    \lower0.6ex\hbox{$\sim$}}}
\def\ltorder{\mathrel{\raise.3ex\hbox{$<$}\mkern-14mu
    \lower0.6ex\hbox{$\sim$}}}

\newcommand{\apj}{ApJ}

\newcommand{\apjl}{ApJL}

\newcommand{\mnras}{MNRAS}
\newcommand{\aj}{AJ}
\newcommand{\araa}{ARA\&A}

\setcitestyle{aysep={},citesep={,}} 

\title[How to Flip a Bar]{How to Flip a Bar} 
\author[A. Collier and A~M. Madigan]
{Angela Collier$^{1}$ and 
Ann-Marie Madigan$^{1}$\thanks{E-mail: annmarie.madigan@colorado.edu}\\
$^{1}$JILA and Department of Astrophysical and Planetary Sciences, CU Boulder, Boulder, CO 80309, USA}

\begin{document}
\date{Accepted ?; Received ??; in original form ???}
\maketitle

\begin{abstract}
Galactic bars, made up of elongated and aligned stellar orbits, can lose angular momentum via resonant torques with dark matter particles in the halo and slow down. 
Here we show that if a stellar bar is decelerated to zero rotation speed, it can flip the sign of its angular momentum and reverse rotation direction.  We demonstrate this in a collisionless $N$-body simulation of a galaxy in a live counter-rotating halo. Reversal begins at small radii and propagates outward. The flip generates a kinematically-decoupled core both in the visible galaxy and in the dark matter halo, and counter-rotation generates a large-scale warp of the outer disk with respect to the bar. 
\end{abstract}


\section{Introduction}
\label{sec:intro}

Galactic bars experience dynamical friction from the dark matter halo \citep{tremaineweinberg84,wein85,athana03,Debattista1998,deb&sell,Chiba2022,Chiba2023}. This friction slows the bar and allows it to grow as increasing numbers of stars enter into resonance \citep{athan92}. Recently, \citet{Chiba2021} demonstrated the deceleration of the Milky Way bar of more than 24\% since its formation to its current pattern speed of $\Omega_b = 35.5 \pm 0.8 ~ \rm{km s}^{-1} \rm{kpc}^{-1}$. Here we ask the question, what happens if a bar is slowed to zero pattern speed? 

To explore this question, we run a numerical experiment of an $N$-body galaxy embedded in a counter-rotating (retrograde) live dark matter halo which acts as a reservoir of negative angular momentum. We have previously shown that a massive stellar bar torques low-inclination retrograde dark matter orbits to prograde orientations, absorbing negative angular momentum in the process \citep{coll19b, CollierandMad20, Lieb2022}. 
These bar-driven orbit reversals act as an effective frictional force on the bar, rapidly decreasing its pattern speed. Here we show that this mechanism can slow the bar to zero pattern speed, flip it in orientation (from the inside-out) and reverse its rotation. This spontaneously creates a kinematically-decoupled core in an isolated galaxy. 
In Section~\ref{sec:ics} we set up the numerical experiment. In Section~\ref{sec:results} we present our results and we conclude in Section~\ref{sec:discussion}. 

\begin{figure*}
    \centering
    \includegraphics[width=\textwidth]{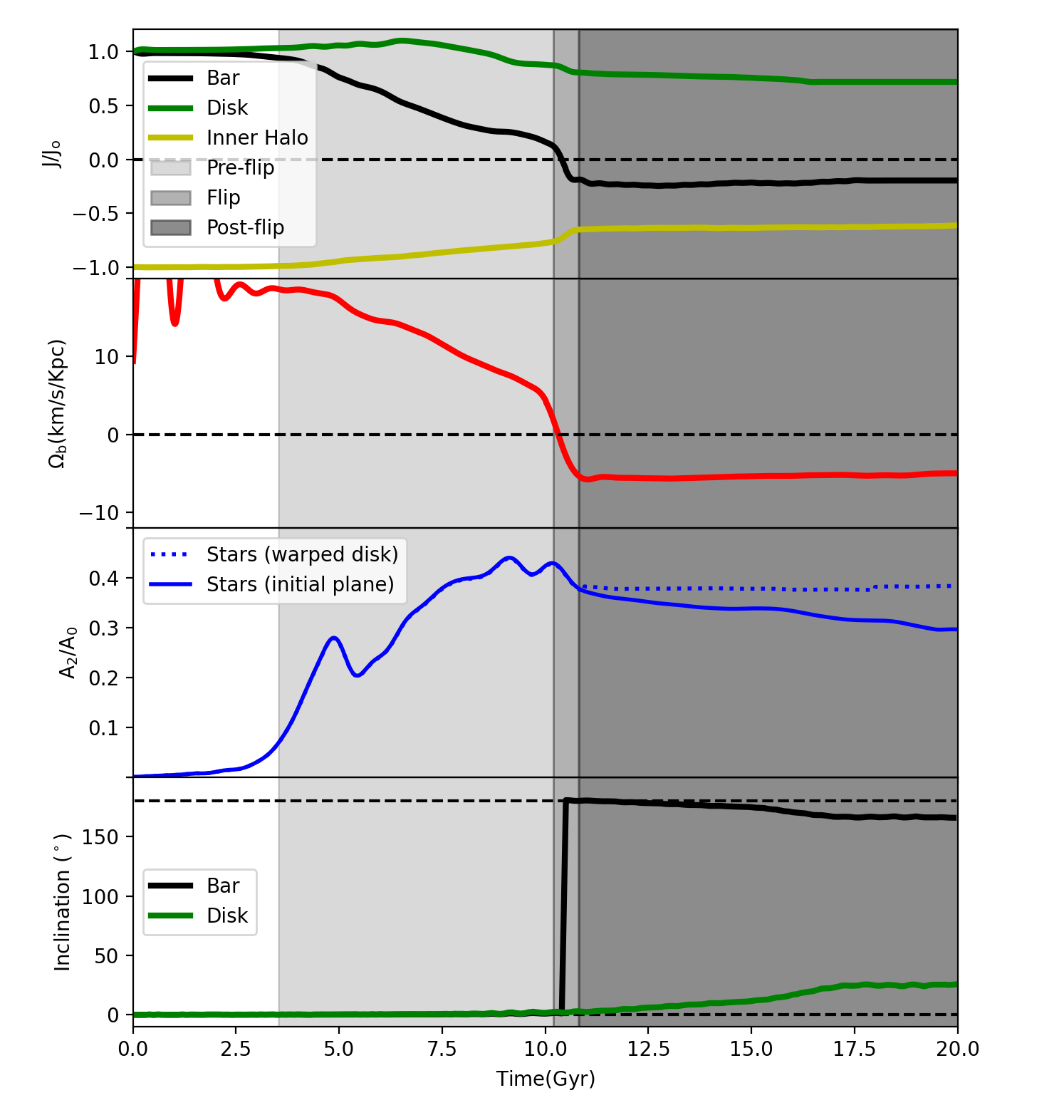}
    \caption{Time evolution of galaxy parameters. 
    The top panel shows the evolution of angular momentum normalized to initial values in three regions: the stellar bar (black), the stellar disk (green) and the inner dark matter halo (yellow). Panel two shows the pattern speed of the bar, $\Omega_b$. Panel three shows the strength of the stellar bar (solid blue line). The disk warps out of the plane at late times; the short-dashed blue line shows the measurement accounting for the warp. The bottom panel shows the inclination of the outer disk and bar with respect to initial plane of the galaxy.
    Background shading denotes the direction of rotation of the stellar bar: light gray while bar is prograde with respect to the disk (pre-flip), darker gray while the bar stalls in rotation (flip), and dark gray while bar is rotating retrograde with respect to the disk (post-flip).}
   \label{fig:sA2}
\end{figure*}

\section{Numerical Simulation Initial Conditions}
\label{sec:ics}

We simulate a stellar disk embedded in a live counter-rotating dark matter halo. The halo is initialized with an NFW-inspired \citep{nava96} density profile,
\begin{equation}
\rho_{\rm h}(r) = \frac{\rho_{\rm s}\,e^{-(r/r_{\rm t})^2}}{[(r+r_{\rm c})/r_{\rm s}](1+r/r_{\rm s})^2},
\end{equation}
where $\rho_{\rm s} \approx7 \times 10^{10}\,M_\odot$/kpc$^3$ is the fitting density parameter, $r_{\rm s}=9$\,kpc is the characteristic radius, and $r_{\rm c}=1.4$\,kpc is a central density core.   The Gaussian cutoff is applied at $r_{\rm t}=86$\,kpc. The dark matter  halo contains $7.2\times 10^6$ particles and the halo mass is $M_{\rm h} = 6.3\times 10^{11}\,M_\odot$.

A non-rotating velocity distribution is generated using an iterative method from \citet{rodio06} (see also \citet{rodio09,coll19b}). 
We reverse the tangential velocities of all prograde particles to create a halo that is retrograde with respect to the stellar disk. The new velocity distribution maintains the solution to the Boltzmann equation and does not alter the velocity profile \citep{lynd60,wein85}, so the equilibrium state is preserved. 
The halo spin parameter is $\lambda \equiv J_h/\sqrt{2}M_{\rm vir}R_{\rm vir}v_c = 0.101$ where $J_h$ is the halo angular momentum, $M_{\rm vir}$ and $R_{\rm vir}$ are the viral mass and radius of the dark matter halo, and $v_c$ is the circular velocity of the system at $R_{\rm vir}$.
The volume density of the exponential stellar disk is
\begin{eqnarray}
\rho_{\rm d}(R,z) = \bigl(\frac{M_{\rm d}}{4\pi h^2 z_0}\bigr)\,{\rm exp}(-R/h)
     \,{\rm sech}^2\bigl(\frac{z}{z_0}\bigr),
\end{eqnarray}
where $M_{\rm d} = 6.3\times 10^{10}\,M_\odot$ is the disk mass, $h=2.85$\,kpc is its radial scale length, and $z_0=0.6$\,kpc is the vertical scale height.  The stellar disk has $N_{\rm d} = 0.8\times 10^6$ particles. The radial and vertical dispersion velocities of stellar particles are given by
\begin{equation}
    \sigma_R(R) = \sigma_{R,0} e^{-R/2h}
\end{equation}
\begin{equation}
    \sigma_z(R) = \sigma_{z,0} e^{-R/2h}
\end{equation}
where $\sigma_{R,0}$ = 120 km/s and $\sigma_{z,0}$ = 100 km/s. 
We evolve the simulation using the $N$-body part of the tree-particle-mesh Smoothed Particle Hydrodynamics (SPH/$N$-body) code GIZMO \citep{hop15}.  The opening angle of the tree code is 0.4 pc and the softening parameter is 0.25 pc.

We note that the initial conditions for this simulation are the same as that presented in \citet{CollierandMad20} and further explored in \citet{Lieb2022}. Here we extend the simulation, long past a Hubble time, to explore the bar flip and subsequent dynamics. Angular momentum and energy are conserved at the level of $8 \times 10^{-4}$ and $7 \times 10^{-4}$ respectively across 20 Gyr.

\section{Results}
\label{sec:results}

Figure~\ref{fig:sA2} shows the evolution of our system. 
We first describe the main results and then break-down the details by stage: pre-flip, flip, and post-flip. 
The initially axisymmetric stellar disk forms a bar after $\sim3.5$ Gyr, the instability being delayed due to the rotation of the dark matter halo \citep{Saha13}.
To quantify the evolution of the bar parameters we first identify bar length ($R_b$). We do this by measuring ellipticity profiles of the isodensity contours in the $x/y$-plane. Bar length is determined by measuring the radius where the ellipticity of the contour has decreased by 15\% from the maximum value \citep{marti06}. The stellar bar reaches a maximum length (of $R_b \sim 14$ kpc) around $11$ Gyr and maintains that length for the remainder of the simulation. We therefore define membership to the stellar bar\footnote{We find the same results if we choose a shorter bar length of $R_b \sim 10$ kpc which is the visual length of the bar in, for example, Figure~\ref{fig:vlos}.} as $R \leq 14$ kpc, the outer disk as $R > 14$ kpc, and the inner dark matter halo as $R \leq 14$ kpc and $|z| \leq 2.5$ kpc.

The top row of Figure~\ref{fig:sA2} shows the evolution of angular momentum of the system in these three distinct regions. Background shading highlights times when the bar angular momentum is aligned with that of the disk (``pre-flip''; light gray), when it appears to stall in its rotation (``flip''; darker gray), and when its angular momentum is opposite that of the disk (``post-flip''; dark gray.)
The second row shows the evolution in pattern speed of the stellar bar, $\Omega_b$. 
The evolution of stellar bar strength is shown in the third row with the solid blue line. The strength of the bar is defined by the ratio of the Fourier $m=2$ mode to the $m=0$ mode,
\begin{eqnarray}
\label{eq:a2}
\frac{A_m}{A_0} = \frac{1}{A_0}\sum_{j=1}^{N_{\rm d}} M_{\rm j}\,e^{mi\phi_{\rm j}},
\end{eqnarray} 
calculated by summing over all disk particles with $R \leq 14$ kpc, and mass $M_j$ at azimuthal angle $\phi_j = \tan^{-1}(y/x)$, and setting $m=2$.  

\begin{figure}
    \centering
\includegraphics[width=0.5\textwidth]{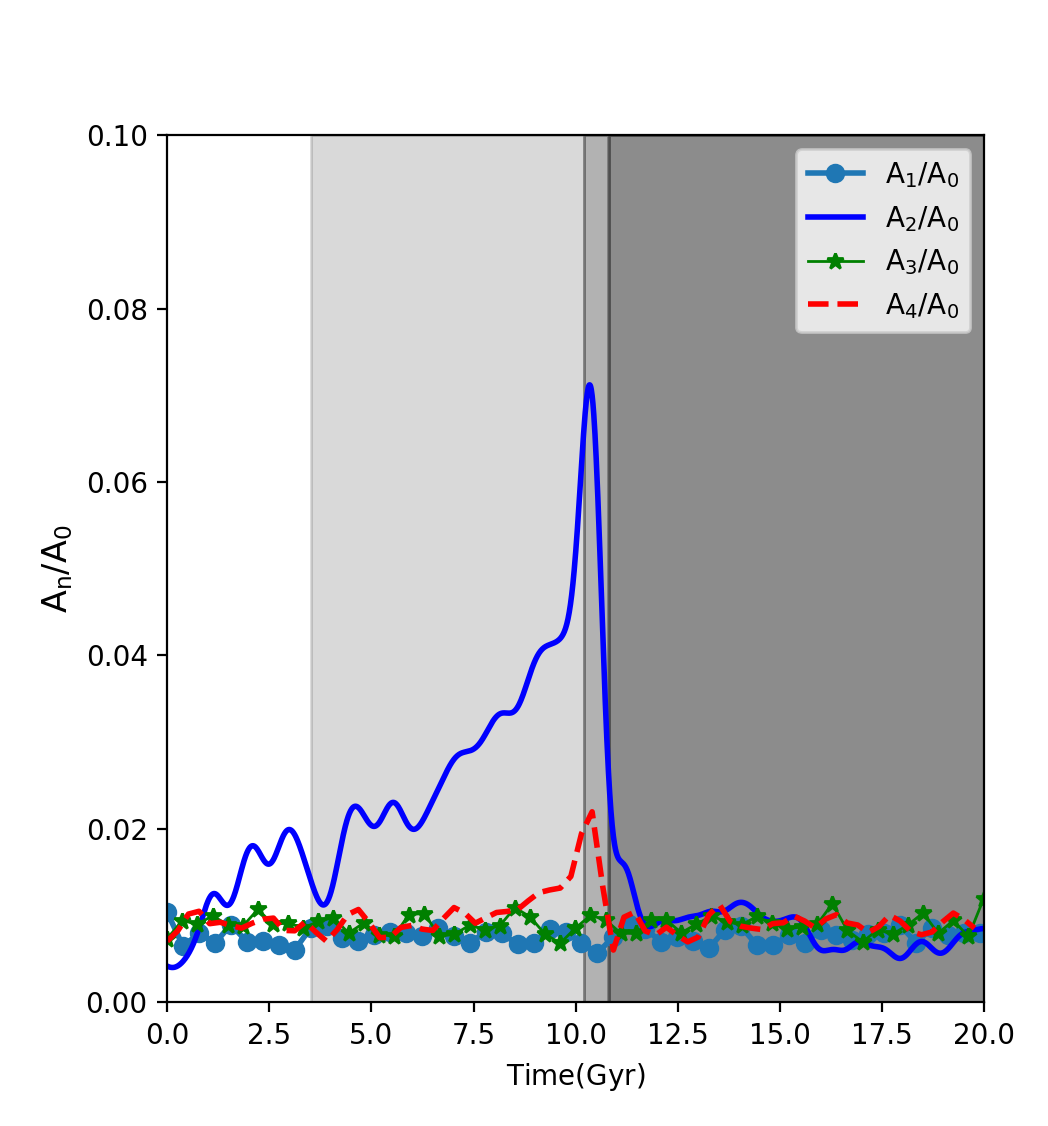}
    \caption{Fourier analysis of the first four modes in the dark matter halo. Note the smaller vertical scale in comparison to the stellar bar mode measurement in Figure~\ref{fig:sA2}. 
    }
   \label{fig:dm_mode}
\end{figure}

We repeat the measurement in equation~\ref{eq:a2} for  $m= 1, 2, 3, 4$ for the inner dark matter halo in Figure~\ref{fig:dm_mode}. These modes quantify the strength of the dark matter response to the presence of the stellar bar. The $m=2$ mode is the strongest; we have previously shown that retrograde halo orbits form a dark matter wake oriented perpendicular to the bar. This disperses when the bar reverses around 10 Gyr. The $m=1$ mode hovers at the noise level for the length of the simulation \citep[cf.][]{Weinberg2022}.

We continue the simulation past a Hubble time to check that the reversed bar is stable and can persist over a long timescale; we confirm that it does. We find that a strong warp develops as the bar begins to counter-rotate within the stellar disk. 
The bottom panel of Figure~\ref{fig:sA2} shows the inclination of the stellar disk rising in response to the bar, and the bar tilting away from an inclination of $180^\circ$. 
Inclination is defined here as $i = \arccos{j_z/|j|}$, where $j$ is the angular momentum and $j_z$ is its $z$-component.
At the end of the simulation the bar is at an inclination of $166^\circ$ and the disk is at an inclination of $26^\circ$, with a separation of $140^\circ$ between the two components.
\newline

    \begin{figure}
    \includegraphics[width=0.5\textwidth]{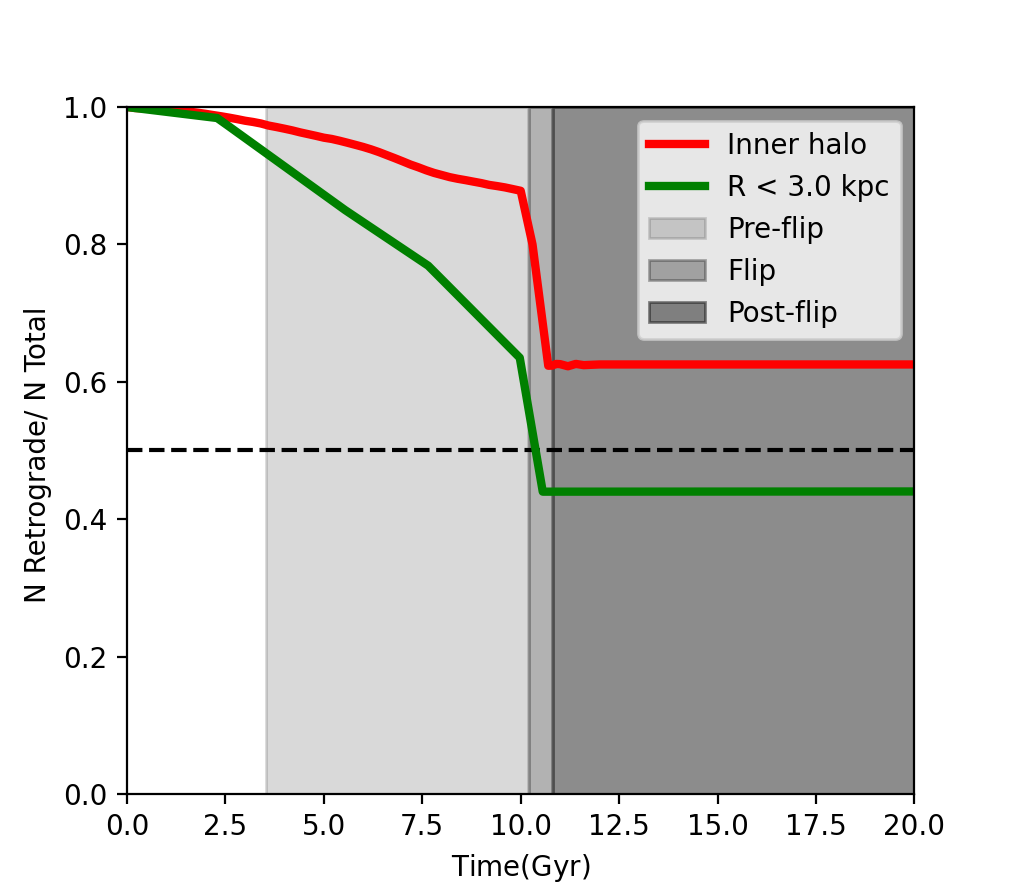}
    \caption{The fraction of dark matter particles of counter-rotating/retrograde orientation as a function of time. Orbit reversals in the inner halo ($R<14$ kpc and $|z|<2.5$ kpc) and inside $R<3$ kpc ($|z|<2.5$ kpc) drive this fraction down especially during the bar flip. The majority of halo orbits with radii less than 3 kpc are prograde after $10$ Gyr resulting in a kinematically decoupled core in the dark matter halo. }
   \label{fig:retplot}
\end{figure}

    \begin{figure}
    \includegraphics[width=0.5\textwidth]{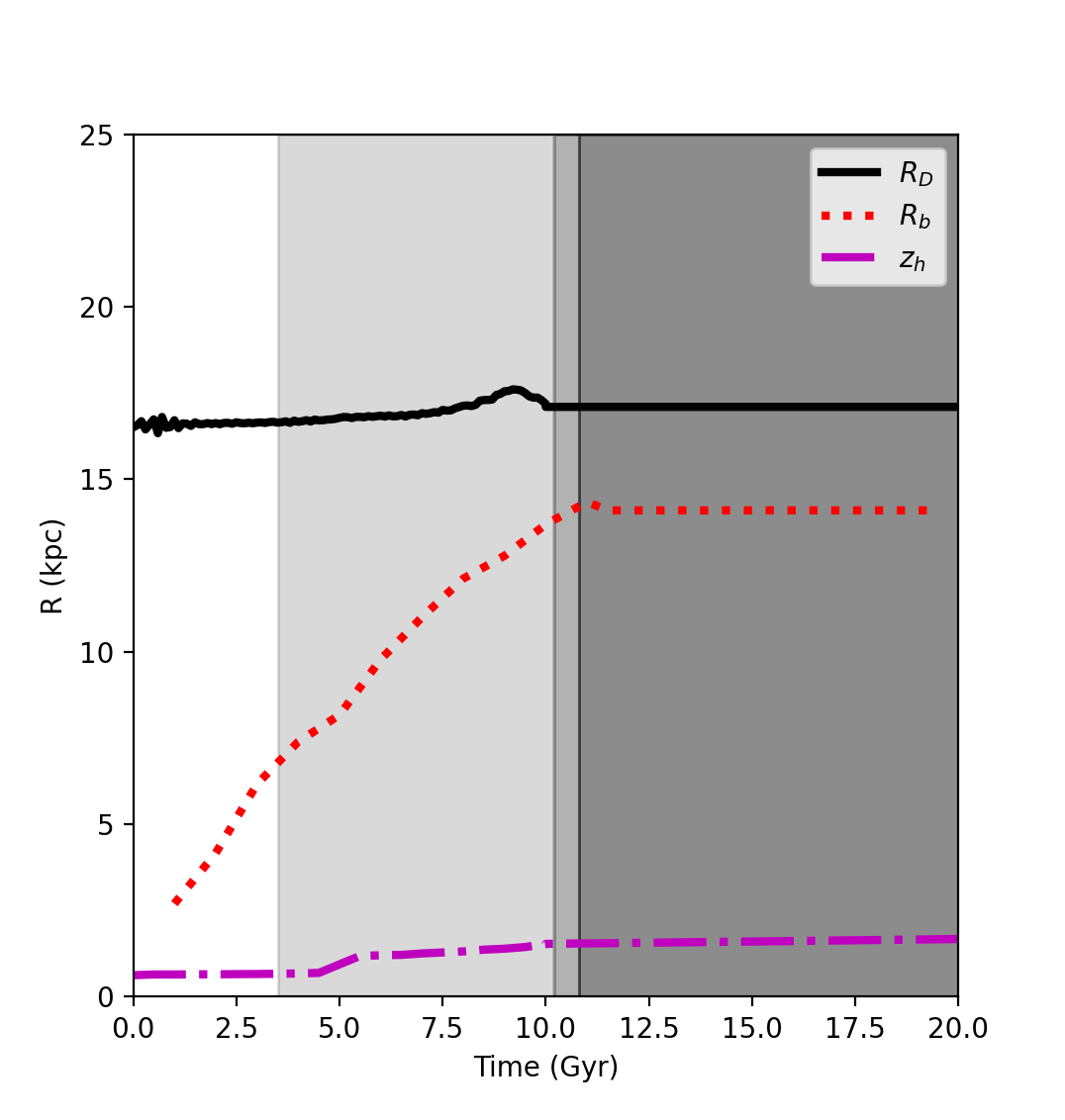}
    \caption{Evolution of disk radius (solid black), bar length (red dashed), and disk scale height (magenta dash dot). These measurements account for the warp of the disk at late times.}
   \label{fig:scale_length}
\end{figure}

    \noindent {\bf Pre-flip, $\mathbf{t \lesssim 10}$ Gyr} \\
    The orbits in the bar lose angular momentum in two main ways. The first is via resonant torques with the stellar disk at corotation and the outer Lindblad resonance. This continues until the bar is sufficiently decelerated such that both radii are pushed beyond the edge of the stellar disk at $\sim7$ Gyr. The second way the bar loses angular momentum is via torques with the inner dark matter halo. This is primarily a result of a population of co-planar dark matter halo orbits reversing orientation from retrograde to prograde via gravitational torques with the aligned stellar orbits of the bar. 
    Bar-driven halo ``orbit reversals'' were first discovered by \citet{coll19b}. A prediction of the proposed explanation given in \citet{CollierandMad20}, that halo orbit reversals are due to the strong gravitational torque of the bar, is that they should occur when coplanar dark matter orbits are oriented ‘behind’ the bar in rotation such that the torque is negative (see their Figure 11 for confirmation).
    \citet{CollierandMad20} show that bar-driven orbit reversals account for more than half of the angular momentum transfer between the stellar bar and the inner dark matter halo (see their Figure 12). 
    Figure \ref{fig:retplot} shows the fraction of retrograde particles in two distinct regions of the dark matter halo. After the bar flip, only $\sim65\%$ of the orbits of the inner dark matter halo are retrograde and within $R<3.0$ kpc the majority of orbits are of prograde orientation. This is due to the  bar-driven orbit reversals. 
    Returning to Figure~\ref{fig:sA2}, over the first $10$ Gyr the bar gets stronger (as measured by $A_2/A_0$), aside from a buckling instability at $\sim 5.5$ Gyr and a weaker buckling-like event at $\sim 9$ Gyr. As angular momentum is carried away from the bar it rapidly decelerates.
    
      In Figure~\ref{fig:scale_length}, we plot 
      time evolution of the scale height of the disk, $z_h$, 
      measured using the ellipse that contains $97\%$ of disk particles. $z_h$ increases from 0.6 kpc to 1.5 kpc across 20 Gyr. The majority of the increase in the thickness of the disk is due to the strong buckling of the elongated bar at $\sim 5.5$ Gyr. We show the same measurement for the disk length, $R_D$, in the mid-plane of the galaxy.
      The disk radius does not expand significantly with time. This is to be expected because the co-rotation radius is rapidly pushed beyond the edge of the disk as the bar decelerates. This prevents the outer disk from resonating and effectively exchanging angular momentum with the bar. The bar radius, $R_b$, increases rapidly until reversal around 10 Gyr. This is consistent with its pattern speed evolution shown in Figure~\ref{fig:sA2}. 
    \newline

    \begin{figure*}
    \centering
    \includegraphics[width=\textwidth]{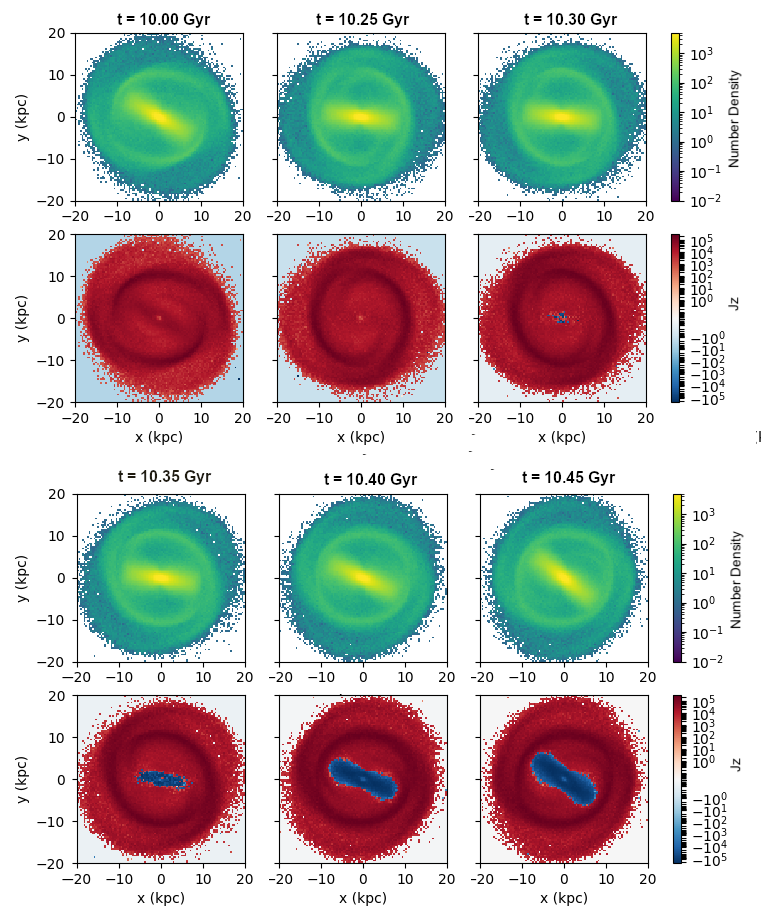}
    \caption{Reversal of a stellar bar. 
    Top: Surface density of stars in bins of length $0.25$ kpc. 
    Bottom: Angular momentum of stars along the line of sight ($J_z$) in units of $M_\odot$ km s$^{-1}$ kpc$^{-1}$. Red indicates particles with positive angular momentum, blue indicates those with negative angular momentum. The stellar bar appears to stall in its counter-clockwise rotation ($10.25-10.35$ Gyr), undergo inside-out reversal ($\gtrsim 10.3$ Gyr), and begin to rotate in the clockwise direction.  
    }
    \label{fig:bar_reverse}
\end{figure*}

\begin{figure*}
    \centering
    \includegraphics[width=\textwidth]{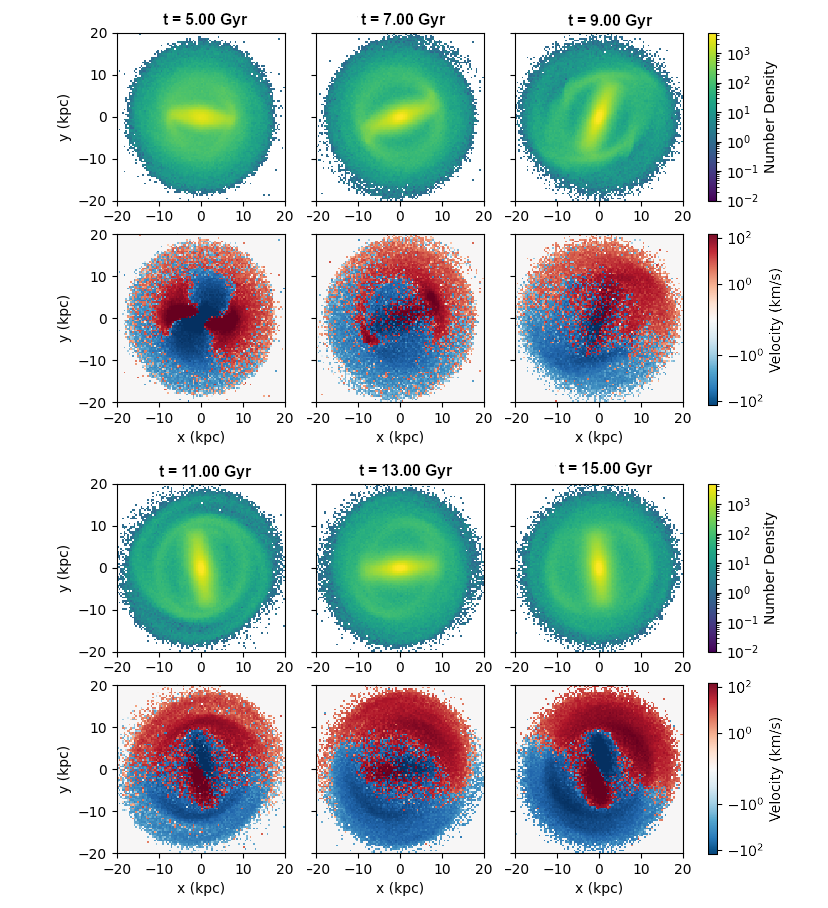}
    \caption{Creation of a kinematically-decoupled core; face-on view. 
    Top: Surface density of stellar particles. 
    Bottom: Line of sight velocity ($v_z$) in $0.25$ pc bins. Red indicates particles moving toward the observer, blue indicates those with moving away. 
   }
   \label{fig:vlos}
\end{figure*}
    
    \noindent {\bf Flip, $\mathbf{t \approx 10}$ Gyr} \\
    At 10.30 Gyr, the bar is torqued through zero angular momentum and flips to negative values for the remainder of the simulation. 
    The steepest reduction in pattern speed occurs while the bar visually appears to stall in its rotation. In Figure~\ref{fig:bar_reverse} we highlight the inside-out reversal of the bar over the 450 Myr that bracket the bar flip. The top rows show the surface density of stars in the disk. The bottom rows show the mean angular momentum in each bin. 
    The reversal begins at small radii and propagates outward.  
    The first frame at $10$ Gyr shows a prograde-rotating (counter-clockwise) bar. 
    Retrograde orbits, shown in blue, begin to dominate the galactic center region at t $\sim 10.3$ Gyr. Smaller bar orbits flip first as they have smaller values of angular momenta. Averaged over the entire stellar bar population, the bar appears to stall for $\mathcal{O}{(100}$) Myr. The final frame shows the bar counter-rotating (clockwise) with a negative pattern speed.
    The pattern speed decreases slightly as it gains a small amount of positive angular momentum (as shown in the top frame of Figure~\ref{fig:sA2}) over the last $8$ Gyr of the simulation.
    The dark matter response is strongest during the flip of the stellar bar, plummeting in strength once the bar reverses. 
    \newline

\begin{figure*}
    \centering
    \includegraphics[width=\textwidth]{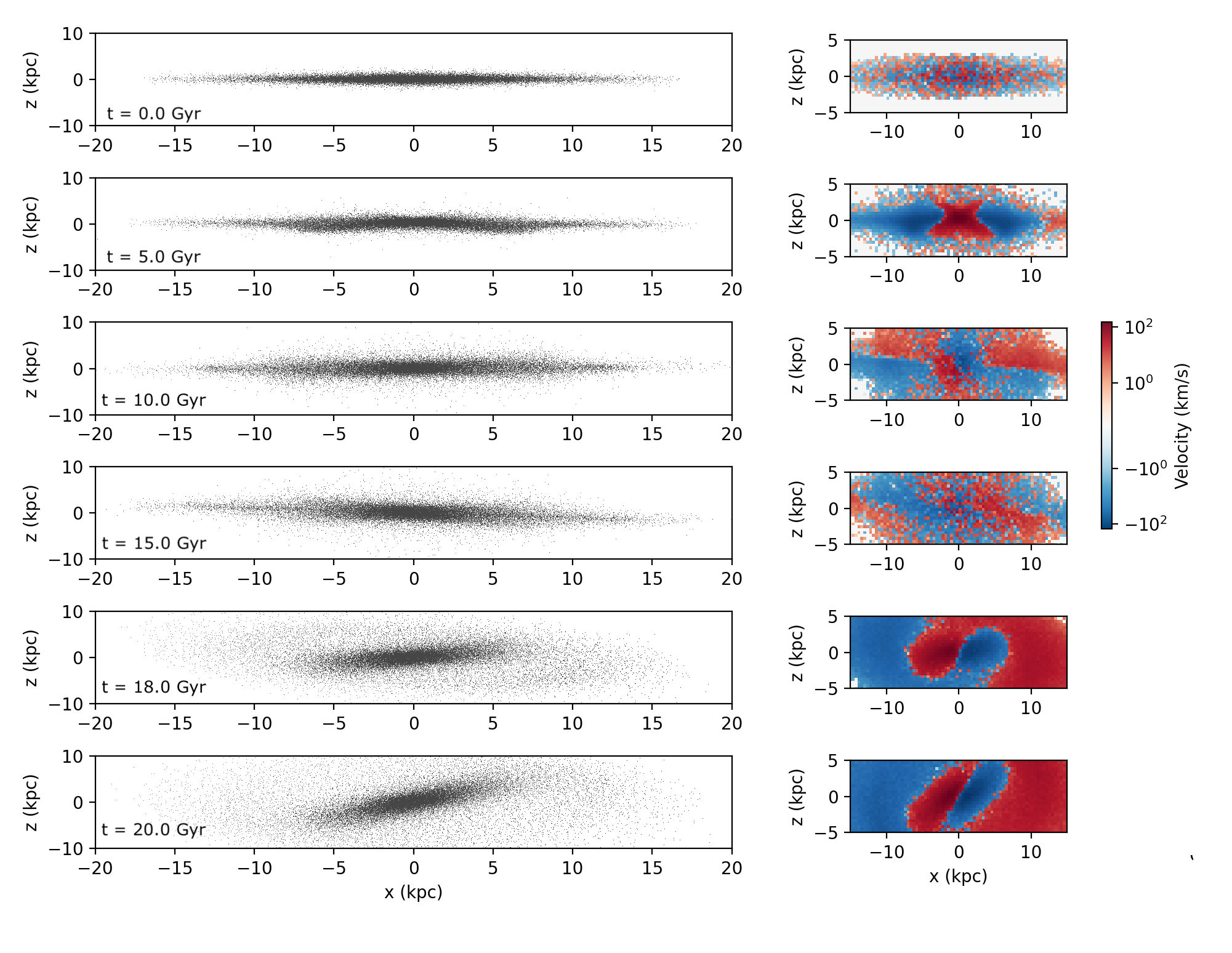}
    \caption{Creation of a kinematically-decoupled core; edge-on view. Left column: Dot plot of disk particles showing the evolution of the initially thin, edge-on disk. Time is indicated in the bottom left of each panel. A strong warp develops at late times. 
   Right column: Line of sight velocity ($v_z$) histogram for a zoomed-in section of the disk midplane at corresponding times.
    }
   \label{fig:disk_title}
\end{figure*}

\begin{figure*}
    \centering
    \includegraphics[width=\textwidth]{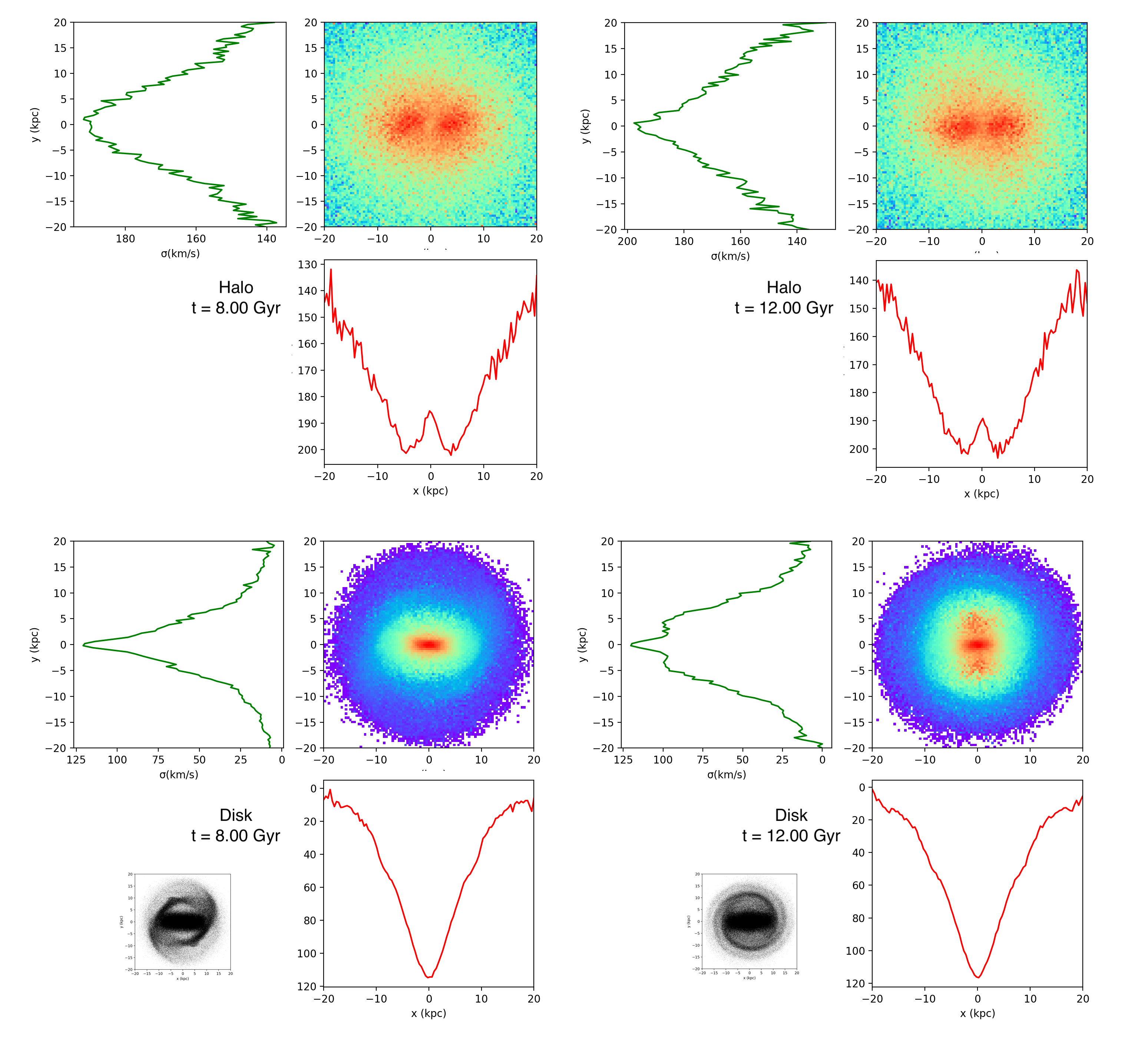}
    \caption{Velocity dispersion along the line-of-sight ($\sigma_z$)  for the halo (top) and disk (bottom) before (left) and after (right) the bar flip. The bar is aligned along the $x-$axis in all frames. The 2D maps show a section of the halo and disk ($|z| \leq 2.5$ kpc) and the 1D histograms show slices along the $x-$axis ($|y| \leq 2.5$ kpc) in red and $y-$axis ($|x| \leq 2.5$ kpc) in green. 
    }
   \label{fig:sigma}
\end{figure*}

    \noindent {\bf Post-flip, $\mathbf{t \gtrsim 10}$ Gyr} \\
    The bar is now kinematically decoupled from the rest of the galaxy. When we account for its tilting plane, the bar stays fixed in strength for the rest of the simulation (10 Gyr). This is seen in the third panel of Figure~\ref{fig:sA2}. 
    Strong spiral arms persist after the bar flips and begins to counter-rotate. The arms trail, and rotate with prograde motion, i.e., in the opposite direction of the bar. 
    The surface density plots in Figure~\ref{fig:bar_reverse} show that an inner ring the same size of the bar appears due to very-tightly wound spiral arms. 
    In Figure \ref{fig:vlos} we plot stellar surface density (top) and line-of-sight velocity histograms (bottom) from 5 \-- 15 Gyr face-on to the galactic disk. This figure showcases the development of a kinematically-decoupled core.
    We run the simulation long past a Hubble time to confirm that the counter-rotating bar persists. We verify that it does and additionally find a large-amplitude warp developing in the galactic plane. We attribute this to the bending instability driven by counter-rotation \citep{toomre64,merritt&sellwood94,Sellwood1994}.

\section{Discussion}
\label{sec:discussion}

In this paper we explore the dynamical evolution of a bar as it is slowed to zero rotation speed. We run a numerical experiment of an N-body galaxy embedded in a live counter-rotating dark matter halo. Co-planar dark matter halo orbits absorb angular momentum from the bar, reversing their orientation. The bar loses angular momentum, slows in response, and flips $180^\circ$ in angular momentum (Figure~\ref{fig:sA2}, bottom panel). This flip spontaneously creates a kinematically-decoupled core in an isolated galaxy (Figure~\ref{fig:vlos}). A large-amplitude warp develops between the counter-rotating bar and outer stellar disk. 

In Figure~\ref{fig:disk_title}, we show edge-on views of the galactic disk with each star represented as a dot (left) and zoomed-in panels in velocity space (right). This figure clearly shows the galactic warp developing at late times alongside the kinematically-decoupled core due to the reversal of the stellar bar. It is also interesting to note the edge-on X-shape in velocity space at 5 Gyr as the bar is buckling \cite[for review see][]{Athanassoula2016}.

Orbit reversals of dark matter halo particles also create a kinematically-decoupled core in the halo (Figure~\ref{fig:retplot}). The dark matter develops an interesting double-peaked velocity dispersion profile along the bar major axis. We show this in Figure~\ref{fig:sigma}, wherein we plot the dispersion values of the velocity along the line of sight, denoted by $\sigma$. The halo is plotted on top; we show plots before (left) and after (right) the bar flip. The bar is aligned along the $x-$axis in all frames. 
The halo hosts a double-peaked $\sigma$ profile aligned with the bar's long axis, both pre- and post-flip. This invites comparison to `double-sigma galaxies' ($2\sigma$) which have two off-centered peaks in their stellar velocity dispersion along the photometric major axis of the galaxy \citep{Krajnovic2011}. A natural explanation for such galaxies is that they host two counter-rotating disks and are thus a subset of multi-spin galaxies \citep{Rubin1994,Corsini2014}. Gas-rich galaxy mergers or external gas accretion and subsequent star formation are most often invoked in explanation \citep[e.g.,][]{Alabi2015}. In our case it is the interaction of retrograde dark matter halo orbits with a massive prograde stellar bar that has generated counter-rotating components in the halo. 
We confirm that this $2\sigma$ feature in the halo is not present in simulations of barred galaxies with an isotropic or prograde-rotating dark matter halo. 

The bottom half of Figure~\ref{fig:sigma} shows the velocity dispersion along the line of sight for the stellar disk. Again the bar is aligned along the $x$-axis both for the pre-flip (left, 8 Gyr) and post-flip distribution (right, 12 Gyr). 
Pre-flip, the velocity dispersion values follow oval-shaped contours aligned with the bar's long axis as expected \citep{Du2017}. Post-flip, there is a central velocity dispersion peak along with two off-centered peaks extending from the bar's short axis where there is a mix of stars moving toward and away from our line of sight. 

Rapid angular momentum changes which can lead to flips in orbital orientation have been studied in near-Keplerian potentials \citep{Katz2012,Antognini2014,Grishin2018,Bhaskar2021}. Notably, numerical results conflict with secular theory when considering extremely low angular momentum orbits.  
Such orbit flips can contribute to the formation of hot Jupiters orbiting retrograde with respect to their host stars' rotation \citep{Li2014a,Naoz2016}, and generate counter-rotating stellar populations in galactic nuclei containing supermassive black holes \citep{madigan2018,Wernke2019ApJ}.
To our knowledge, collective orbit flips on galactic scales involving hundreds of thousands of particles have not previously been demonstrated. We note however that our collisionless simulation does not include gas dynamics which has been demonstrated to limit bar growth and deceleration \citep{Athanassoula2013,Villa-Vargas2010,Beane2022}.

\section*{Acknowledgements}
AC was supported by an NSF Astronomy and Astrophysics Postdoctoral Fellowship under award AST-2102185.
AM gratefully acknowledges support from the David and Lucile Packard Foundation. This work utilized resources from the University of Colorado Boulder Research Computing Group, which is supported by the National Science Foundation (awards ACI-1532235 and ACI-1532236), the University of Colorado Boulder, and Colorado State University. 

\section*{Data Availability}
The data underlying this article will be shared on reasonable request to the corresponding author.

\bibliographystyle{mn}

\begin{thebibliography}{40}
\expandafter\ifx\csname natexlab\endcsname\relax\def\natexlab#1{#1}\fi

\bibitem[{{Alabi} {et~al.}(2015){Alabi}, {Foster}, {Forbes}, {Romanowsky},
  {Pastorello}, {Brodie}, {Spitler}, {Strader}, \& {Usher}}]{Alabi2015}
{Alabi} A.~B., {Foster} C., {Forbes} D.~A., {Romanowsky} A.~J., {Pastorello}
  N., {Brodie} J.~P., {Spitler} L.~R., {Strader} J., {Usher} C., 2015, \mnras,
  452, 2208

\bibitem[{{Antognini} {et~al.}(2014){Antognini}, {Shappee}, {Thompson}, \&
  {Amaro-Seoane}}]{Antognini2014}
{Antognini} J.~M., {Shappee} B.~J., {Thompson} T.~A., {Amaro-Seoane} P., 2014,
  \mnras, 439, 1079

\bibitem[{{Athanassoula}(1992)}]{athan92}
{Athanassoula} E., 1992, \mnras, 259, 345

\bibitem[{{Athanassoula}(2003)}]{athana03}
---, 2003, \mnras, 341, 1179

\bibitem[{{Athanassoula}(2016)}]{Athanassoula2016}
---, 2016, in Astrophysics and Space Science Library, Vol. 418, Galactic
  Bulges, {Laurikainen} E., {Peletier} R., {Gadotti} D., eds., p. 391

\bibitem[{{Athanassoula} {et~al.}(2013){Athanassoula}, {Machado}, \&
  {Rodionov}}]{Athanassoula2013}
{Athanassoula} E., {Machado} R. E.~G., {Rodionov} S.~A., 2013, \mnras, 429,
  1949

\bibitem[{{Beane} {et~al.}(2022){Beane}, {Hernquist}, {D'Onghia}, {Marinacci},
  {Conroy}, {Qi}, {Sales}, {Torrey}, \& {Vogelsberger}}]{Beane2022}
{Beane} A., {Hernquist} L., {D'Onghia} E., {Marinacci} F., {Conroy} C., {Qi}
  J., {Sales} L.~V., {Torrey} P., {Vogelsberger} M., 2022, arXiv e-prints,
  arXiv:2209.03364

\bibitem[{{Bhaskar} {et~al.}(2021){Bhaskar}, {Li}, {Hadden}, {Payne}, \&
  {Holman}}]{Bhaskar2021}
{Bhaskar} H., {Li} G., {Hadden} S., {Payne} M.~J., {Holman} M.~J., 2021, \aj,
  161, 48

\bibitem[{{Chiba}(2023)}]{Chiba2023}
{Chiba} R., 2023, arXiv e-prints, arXiv:2305.00022

\bibitem[{{Chiba} \& {Sch{\"o}nrich}(2021)}]{Chiba2021}
{Chiba} R., {Sch{\"o}nrich} R., 2021, \mnras, 505, 2412

\bibitem[{{Chiba} \& {Sch{\"o}nrich}(2022)}]{Chiba2022}
---, 2022, \mnras, 513, 768

\bibitem[{{Collier} \& {Madigan}(2021)}]{CollierandMad20}
{Collier} A., {Madigan} A.-M., 2021, \apj, 915, 23

\bibitem[{{Collier} {et~al.}(2019){Collier}, {Shlosman}, \& {Heller}}]{coll19b}
{Collier} A., {Shlosman} I., {Heller} C., 2019, \mnras, 489, 3102

\bibitem[{{Corsini}(2014)}]{Corsini2014}
{Corsini} E.~M., 2014, in Astronomical Society of the Pacific Conference
  Series, Vol. 486, Multi-Spin Galaxies, {Iodice} E., {Corsini} E.~M., eds.,
  p.~51

\bibitem[{{Debattista} \& {Sellwood}(1998)}]{Debattista1998}
{Debattista} V.~P., {Sellwood} J.~A., 1998, \apjl, 493, L5

\bibitem[{{Debattista} \& {Sellwood}(2000)}]{deb&sell}
---, 2000, \apj, 543, 704

\bibitem[{{Du} {et~al.}(2017){Du}, {Shen}, {Debattista}, \& {de
  Lorenzo-C{\'a}ceres}}]{Du2017}
{Du} M., {Shen} J., {Debattista} V.~P., {de Lorenzo-C{\'a}ceres} A., 2017,
  \apj, 836, 181

\bibitem[{{Grishin} {et~al.}(2018){Grishin}, {Perets}, \&
  {Fragione}}]{Grishin2018}
{Grishin} E., {Perets} H.~B., {Fragione} G., 2018, \mnras, 481, 4907

\bibitem[{{Hopkins}(2015)}]{hop15}
{Hopkins} P.~F., 2015, \mnras, 450, 53

\bibitem[{{Katz} \& {Dong}(2012)}]{Katz2012}
{Katz} B., {Dong} S., 2012, arXiv e-prints, arXiv:1211.4584

\bibitem[{{Krajnovi{\'c}} {et~al.}(2011){Krajnovi{\'c}}, {Emsellem},
  {Cappellari}, {Alatalo}, {Blitz}, {Bois}, {Bournaud}, {Bureau}, {Davies},
  {Davis}, {de Zeeuw}, {Khochfar}, {Kuntschner}, {Lablanche}, {McDermid},
  {Morganti}, {Naab}, {Oosterloo}, {Sarzi}, {Scott}, {Serra}, {Weijmans}, \&
  {Young}}]{Krajnovic2011}
{Krajnovi{\'c}} D., {Emsellem} E., {Cappellari} M., {Alatalo} K., {Blitz} L.,
  {Bois} M., {Bournaud} F., {Bureau} M., {Davies} R.~L., {Davis} T.~A., {de
  Zeeuw} P.~T., {Khochfar} S., {Kuntschner} H., {Lablanche} P.-Y., {McDermid}
  R.~M., {Morganti} R., {Naab} T., {Oosterloo} T., {Sarzi} M., {Scott} N.,
  {Serra} P., {Weijmans} A.-M., {Young} L.~M., 2011, \mnras, 414, 2923

\bibitem[{{Li} {et~al.}(2014){Li}, {Naoz}, {Kocsis}, \& {Loeb}}]{Li2014a}
{Li} G., {Naoz} S., {Kocsis} B., {Loeb} A., 2014, \apj, 785, 116

\bibitem[{{Lieb} {et~al.}(2022){Lieb}, {Collier}, \& {Madigan}}]{Lieb2022}
{Lieb} E., {Collier} A., {Madigan} A.-M., 2022, \mnras, 509, 685

\bibitem[{{Lynden-Bell}(1960)}]{lynd60}
{Lynden-Bell} D., 1960, \mnras, 120, 204

\bibitem[{{Madigan} {et~al.}(2018){Madigan}, {Halle}, {Moody}, {McCourt},
  {Nixon}, \& {Wernke}}]{madigan2018}
{Madigan} A.-M., {Halle} A., {Moody} M., {McCourt} M., {Nixon} C., {Wernke} H.,
  2018, \apj, 853, 141

\bibitem[{{Martinez-Valpuesta} {et~al.}(2006){Martinez-Valpuesta}, {Shlosman},
  \& {Heller}}]{marti06}
{Martinez-Valpuesta} I., {Shlosman} I., {Heller} C., 2006, \apj, 637, 214

\bibitem[{{Merritt} \& {Sellwood}(1994)}]{merritt&sellwood94}
{Merritt} D., {Sellwood} J.~A., 1994, \apj, 425, 551

\bibitem[{{Naoz}(2016)}]{Naoz2016}
{Naoz} S., 2016, \araa, 54, 441

\bibitem[{{Navarro} {et~al.}(1996){Navarro}, {Frenk}, \& {White}}]{nava96}
{Navarro} J.~F., {Frenk} C.~S., {White} S. D.~M., 1996, \apj, 462, 563

\bibitem[{{Rodionov} {et~al.}(2009){Rodionov}, {Athanassoula}, \&
  {Sotnikova}}]{rodio09}
{Rodionov} S.~A., {Athanassoula} E., {Sotnikova} N.~Y., 2009, \mnras, 392, 904

\bibitem[{{Rodionov} \& {Sotnikova}(2006)}]{rodio06}
{Rodionov} S.~A., {Sotnikova} N.~Y., 2006, Astronomy Reports, 50, 983

\bibitem[{{Rubin}(1994)}]{Rubin1994}
{Rubin} V.~C., 1994, \aj, 108, 456

\bibitem[{{Saha} \& {Naab}(2013)}]{Saha13}
{Saha} K., {Naab} T., 2013, \mnras, 434, 1287

\bibitem[{{Sellwood} \& {Merritt}(1994)}]{Sellwood1994}
{Sellwood} J.~A., {Merritt} D., 1994, \apj, 425, 530

\bibitem[{{Toomre}(1964)}]{toomre64}
{Toomre} A., 1964, \apj, 139, 1217

\bibitem[{{Tremaine} \& {Weinberg}(1984)}]{tremaineweinberg84}
{Tremaine} S., {Weinberg} M.~D., 1984, \mnras, 209, 729

\bibitem[{{Villa-Vargas} {et~al.}(2010){Villa-Vargas}, {Shlosman}, \&
  {Heller}}]{Villa-Vargas2010}
{Villa-Vargas} J., {Shlosman} I., {Heller} C., 2010, \apj, 719, 1470

\bibitem[{{Weinberg}(1985)}]{wein85}
{Weinberg} M.~D., 1985, \mnras, 213, 451

\bibitem[{{Weinberg}(2022)}]{Weinberg2022}
---, 2022, arXiv e-prints, arXiv:2209.06846

\bibitem[{{Wernke} \& {Madigan}(2019)}]{Wernke2019ApJ}
{Wernke} H.~N., {Madigan} A.-M., 2019, \apj, 880, 42

\end{thebibliography}

\appendix

\end{document}